\documentclass{aa}

\usepackage{graphicx}
\usepackage{amsmath}
\usepackage{amssymb}
\usepackage{verbatim}
\usepackage{array}
\usepackage{txfonts}
\usepackage[switch]{lineno}
\usepackage{multirow}
\usepackage{float}
\usepackage{color}
\usepackage{xspace}  
\usepackage[dvipsnames]{xcolor} 
\usepackage[acronym,nomain]{glossaries}
\usepackage{ulem}
\usepackage[colorlinks=true,linkcolor=blue,citecolor=blue,urlcolor=blue]{hyperref}

\newcommand{\kms}{\,km\,s$^{-1}$\xspace}
\newcommand{\punit}{\,erg\,s$^{-1}$\xspace}

\newcommand{\eunit}{\,erg\xspace}

\newcommand{\kev}{\,keV\xspace}

\newcommand{\mug}{\,$\mu$G\xspace}
\newcommand{\yr}{\,yr\xspace}
\newcommand{\kyr}{\,kyr\xspace}
\newcommand{\pc}{\,pc\xspace}
\newcommand{\kpc}{\,kpc\xspace}
\newcommand{\ghz}{\,GHz\xspace}
\newcommand{\mjy}{\,mJy\xspace}

\newcommand{\ujy}{\,$\mu$Jy\xspace}
\def\deg{\ensuremath{^\circ}\xspace}

\newacronym{ccsne}{ccSNe}{core-collapse supernovae}
\newacronym{cd}{CD}{contact discontinuity}
\newacronym{cmb}{CMB}{Cosmic Microwave Background}
\newacronym{cta}{CTA}{Cherenkov Telescope Array}
\newacronym{cr}{CR}{cosmic-ray}
\newacronym{crs}{CRs}{cosmic rays}
\newacronym{gc}{GC}{Galactic Center}
\newacronym{gps}{GPS}{Galactic Plane Survey}
\newacronym{fs}{FS}{forward shock}
\newacronym{he}{HE}{High Energy}
\newacronym{hgps}{HGPS}{H.E.S.S. Galactic plane survey}
\newacronym{iact}{IACT}{Imaging Atmospheric Cherenkov Telescope}
\newacronym{ic}{IC}{Inverse Compton}
\newacronym{ism}{ISM}{interstellar medium}
\newacronym{isrf}{ISRF}{interstellar radiation field}
\newacronym{lmc}{LMC}{Large Magellanic Cloud}
\newacronym{mw}{MW}{Milky Way}
\newacronym{psr}{PSR}{pulsar}
\newacronym[plural=PWNe,firstplural=pulsar wind nebulae (PWNe)]{pwn}{PWN}{pulsar wind nebula}
\newacronym{rs}{RS}{reverse shock}
\newacronym{sdr}{SDR}{suppressed diffusion region}
\newacronym{sn}{SN}{supernova}
\newacronym{sne}{SNe}{supernovae}
\newacronym{spp}{SPP}{SNR-PSR-PWN}
\newacronym{ts}{TS}{termination shock}
\newacronym[plural=SNRs,firstplural=supernova remnants (SNRs)]{snr}{SNR}{supernova remnant}
\newacronym{uhe}{UHE}{ultra-high-energy}
\newacronym{vhe}{VHE}{very-high-energy}

\newcommand\vpsr{V_{\mathrm{PSR}}}

\begin{document} 

\title{Radio streaks in the Lighthouse Nebula discovered with MeerKAT}
\subtitle{Particles escaping from the tail and illuminating the ambient magnetic field}

\author{Pierrick Martin\inst{\ref{irap}}\thanks{pierrick.martin@irap.omp.eu}  \and
	    Mickael Coriat\inst{\ref{irap}} \and
	    Barbara Olmi\inst{\ref{oaa}} \and
	    Elena Amato\inst{\ref{oaa},\ref{unifi}} \and
	    Niccol\`o Bucciantini\inst{\ref{oaa},\ref{unifi},\ref{infn}} \and \\
	    Sanja Lazarevi\'c\inst{\ref{wsu},\ref{atnf},\ref{aob}} \and
	    Miroslav D. Filipovi\'c\inst{\ref{wsu}} \and
	    Alexandre Marcowith\inst{\ref{lupm}} \and
	    Sarah Recchia\inst{\ref{oaa},\ref{ifj}}}
        
\authorrunning{Martin et al.}

\institute{
IRAP, Universit\'e de Toulouse, CNRS, CNES, F-31028 Toulouse, France \label{irap} \and
Osservatorio Astrofisico di Arcetri, INAF, Largo Enrico Fermi 5 I-50125, Firenze, Italy \label{oaa} \and
Università degli Studi di Firenze, Via Sansone 1 I-50019, Sesto Fiorentino (FI), Italy \label{unifi} \and 
\label{infn} INFN , Sezione di Firenze, Via G. Sansone 1, I-50019 Sesto Fiorentino (FI), Italy \and
Western Sydney University, Locked Bag 1797, Penrith South DC, NSW 2751, Australia \label{wsu} \and
Australia Telescope National Facility, CSIRO, Space and Astronomy, PO Box 76, Epping, NSW 1710, Australia \label{atnf} \and
Astronomical Observatory, Volgina 7, 11060 Belgrade, Serbia \label{aob} \and
LUPM, Universit\'e de Montpellier, CNRS/IN2P3, CC72, Place Eug\`ene Bataillon, F-34095, France \label{lupm} \and
IFJ-PAN, Institute of Nuclear Physics Polish Academy of Sciences, PL-31342 Krakow, Poland \label{ifj}
}

\date{Received XXX; Accepted YYY}

  \abstract
{Bow-shock pulsar wind nebulae are valuable sources for investigating the dynamics of relativistic pulsar winds and the mechanisms by which they produce cosmic-ray leptons at the highest energies. The Lighthouse Nebula is one such object, famous for the high velocity of its pulsar and its long, misaligned X-ray jet, which may provide a specific escape channel for the most energetic particles.}
{We aim to understand how the bulk of non-thermal particles is released into the interstellar medium. We focus on gigahertz radio observations, which probe lower-energy particles that are dominant in number and long-lived, thus providing a picture of how escape proceeds over long timescales.}
{We analysed MeerKAT and ASKAP observations at 1.28\ghz and 0.94\ghz, respectively.}
{MeerKAT observations reveal a highly structured synchrotron nebula downstream of the pulsar PSR~J1101--6101. A cometary tail extends beyond 5\pc from the pulsar, while we detect multiple two-sided transverse emission streaks for the first time. We detect no radio counterpart of the misaligned X-ray jet, but the spectral index map shows an apparent asymmetry, with the steepest values near the base of the jet.
}
{We interpret the radio streaks as the occasional, charge-independent release of energetic leptons from the tail into the surrounding medium, resulting from dynamical instabilities and reconfiguration in the downstream flow. Most of the particle content of the nebula is discharged into the ambient medium within several parsecs. Escaped particles light up the ambient magnetic field, which appears to have a coherence length of at least a few parsecs. The length and persistence of the streaks indicate a low level of magnetic turbulence, possibly slightly enhanced with respect to average cosmic-ray transport conditions in the Galaxy. This may result from self-generated turbulence caused by a resonant streaming instability or from past activity of the progenitor star.}

\keywords{pulsars:general -- Acceleration of particles -- ISM: magnetic fields -- ISM: individual objects: Lighthouse PWN}

\maketitle

\section{Introduction}

Pulsars are rapidly spinning and highly magnetised neutron stars formed in most core-collapse supernovae. They store about $10^{48}-10^{49}$\eunit of rotational energy, which they return to their surroundings over $10^{2}-10^{5}$\yr timescales. Pulsars release their energy by producing a cold and magnetised pair-plasma outflow with Lorentz factors of $10^{4}-10^{7}$ (the presence of ions in this wind is still debated). The interaction of this ultra-relativistic wind with the surrounding medium generates a shock where the kinetic energy of the wind is converted into non-thermal particles. Although not yet fully elucidated, the process is inferred to be very efficient (with conversion rates of $\sim$$10-100$\%) and to produce very extended particle distributions (ranging from gigaelectronvolts  to petaelectronvolts). The structure that forms downstream of the wind termination shock (TS), up to the contact discontinuity (CD), separating the shocked pulsar wind and stellar ejecta, is called a pulsar wind nebula (PWN). It essentially consists of non-thermal electron-positron pairs bathed in a strong and turbulent magnetic field. As a result, PWNe shine over the entire electromagnetic spectrum through the production of intense synchrotron and inverse-Compton scattering radiation and constitute a prominent class of Galactic sources, especially in the radio, X-ray, and gamma-ray domains. ~\citet{Mitchell:2022,Olmi:2023a,Olmi:2023b} provide recent reviews of PWNe.

During its early evolution~($t \lesssim 10^3$\yr), the PWN is fed by a maximally powerful pulsar and grows rapidly within cold stellar ejecta~\citep[$R_{\rm PWN}\propto t^{6/5}$;][]{VanDerSwaluw:2001}. At intermediate ages~($10^4 \lesssim t \lesssim 10^5$\yr), the PWN evolution is instead determined by its interaction with the reverse shock~(RS) of the supernova remnant~(SNR), an inverse shock moving from the outer boundary of the ejecta towards the centre of the stellar explosion, compressing and heating the stellar ejecta. 
The PWN ends up confined in a high-pressure medium and, depending on its energetic reservoir, might compress and re-expand under the pressure exerted by the shocked ejecta (one-zone models actually predict multiple cycles of compression and expansion that are likely suppressed in three dimensions; see e.g.~\citealt{Bandiera:2023} and the references therein), modifying its morphology and spectral properties. The PWN growth is shallower at late times~\citep[$R_{\rm PWN}\propto t^{3/10}$;][]{VanDerSwaluw:2001}. In this stage, the pulsar can escape the PWN and then the SNR as a result of the initial impulse it received in the supernova explosion. The observed proper motions imply typical velocities of $100-1000$\kms \citep{Verbunt:2017,Igoshev:2020}, suggesting that the pulsar typically escapes into the interstellar medium (ISM) about $50-100$\kyr after the explosion ~\citep[see][for a revision]{2026AA...706A.140B}.

The interaction of the pulsar with the ISM leads to the creation of a specific subclass of objects called bow-shock pulsar wind nebulae (BSPWNe), in which the nebula is confined by the ram pressure of the medium crossed by the pulsar~\citep{Gaensler:2006,Bykov:2017}. High-resolution X-ray observations of the synchrotron emission from BSPWNe, especially those obtained with the Chandra X-ray Observatory, have been key to revealing the nature of these objects and their remarkable diversity~\citep{Kargaltsev:2017}. Valuable complementary information comes from measurements in the radio domain~\citep{Kothes:2017}. Particles emitting at gigahertz frequencies have long radiative lifetimes and should provide a time-integrated picture of the particle population. Moreover, their smaller Larmor radii make them sensitive to different acceleration and transport processes. Line observations in the optical and UV bands are also relevant probes of BSPWNe in cases where the pulsar moves through a partially ionised medium and collisional ionisation or charge exchange can occur in the bow-shock head or tail. 

Approximately three dozen firm or candidate BSPWNe are known, and none appear to be alike~\citep{Kargaltsev:2017}. They exhibit features as diverse as bipolar jets, torus-like components, bow-shaped shock fronts, long or short cometary tails, and one- or two-sided misaligned coherent jets. The physical setup leading to the formation of all these structures, or a specific combination thereof, remains unaccounted for. It likely depends on numerous parameters intrinsic to the pulsar or related to the state of the ambient medium and requires a multi-scale, multi-physics description, from AU-scale charged-particle acceleration and transport to parsec-scale supersonic and turbulent fluid dynamics. The recent discovery of extended gamma-ray halos around some of these objects has renewed interest in the subject, although it is still unclear how common this phenomenon is or whether it is connected to the pulsar itself or the ambient medium~\citep{LopezCoto:2022,Amato:2024b}. Overall, BSPWNe are fascinating astrophysical systems that provide insight into a rich variety of topics: the structure of pulsar magnetospheres and PWNe; particle acceleration mechanisms in relativistic and turbulent flows; particle escape from a source, environmental conditions in the ISM, and early-stage non-thermal particle transport, with a connection to the question of direct measurements of local cosmic-ray leptons.

Among BSPWNe, supersonically moving pulsars form a specific subset of neutron stars with kick velocities $\gtrsim 300$\kms, well above the sound speed in the cold and warm phases of the ISM at $3-30$\kms~\citep{Kargaltsev:2017}. This configuration is expected to produce characteristic morphological patterns for the TS and CD~\citep{Bucciantini:2005,Vigelius:2007}, as well as relatively focused, extended tails. Prominent members of the class are PSR~B2224+65 and the Guitar Nebula~\citep{Hui:2007a}, PSR~J1101–6101 and the Lighthouse Nebula~\citep{Pavan:2016}, PSR~J2030+4415~\citep{DeVries:2022}, PSR~J1509--5850~\citep{Hui:2007b}, and the recently discovered PSR J1638--4713 and the Potoroo Nebula~\citep{Lazarevic:2024}. However, unambiguous identification of actual BSPWNe supersonically moving in the ISM is challenging, and distinguishing them from younger PWNe still residing in their parent SNR and disrupted by a strongly asymmetrical reverse shock is not straightforward.

Interestingly, about half a dozen supersonic BSPWNe exhibit so-called misaligned filaments~\citep{Dinsmore:2024}. These extend over parsec scales in the ISM with a linear, narrow morphology, in an oblique direction with respect to the pulsar's proper motion. In a few cases, a weaker, shorter anti-filament is observed. They appear exclusively in X-rays and are thought to consist of synchrotron-emitting, high-energy pairs that escape along ambient field lines from a specific location near the pulsar bow-shock apex. \citet{Olmi:2024} discuss the theoretical challenges posed by these features and attempt to account for them through non-resonant streaming instability.

No radio counterpart of X-ray misaligned filaments has been observed so far, suggesting that the escape process might require an energy threshold that radio-emitting particles do not reach. A common view on the phenomenon is that only the highest-energy pairs can flow across the specific magnetic configuration formed when the bow-shock structure reconnects with the ambient field on one side of the head, while lower-energy particles are advected downstream. This fits with the fact that BSPWNe tails have been observed in the radio. Simulations of BSPWNe illustrate the asymmetric and energy-dependent nature of particle escape \citep{Barkov:2019a,Olmi:2019}, and this method has been successfully applied to observations of, for example, PSR J2030+4415~\citep{DeVries:2022}.

Despite the non-detection of misaligned jets in the radio~\citep[apart perhaps from that reported in][in the Spaghetti Nebula]{Khabibullin:2024}, long non-thermal filaments observed in the Galactic centre region have been proposed as low-frequency analogues of misaligned X-ray jets. These radio filaments would be magnetic flux tubes that are currently illuminated after having been connected to fast-moving pulsars in the recent past~\citep{Barkov:2019b}. X-ray observations of the BSPWN candidate G0.13--0.11, located ahead of an array of bright non-thermal filaments, support this scenario, revealing misaligned jets and a polarimetric signal consistent with the expected properties~\citep{Churazov:2024}. Non-thermal radio filaments are valuable probes of particle transport processes at gigaelectronvolt particle energies, where most of the cosmic-ray power lies~\citep{Thomas:2020}. 

In this paper, we focus on MeerKAT and ASKAP (Australian Square Kilometre Array Pathfinder) radio observations of a region encompassing PSR J1101--6101 and the Lighthouse Nebula, together with the SNR MSH 11--61A. The Lighthouse Nebula was originally detected as a hard X-ray source IGR J11014--6103, located south-west of the SNR MSH 11--61A, and exhibits a complex X-ray morphology that comprises a point source, an elongated cometary tail, a prominent one-sided jet-like feature perpendicular to it, and a weaker, smaller counter-jet~\citep{Pavan:2011,Pavan:2014,Pavan:2016}. The detection of PSR~J1101--6101, with a period $P = 62.8$\,ms, a spin-down power of $L = 1.36 \times 10^{36}$\punit, and a characteristic age of 116\kyr~\citep{Halpern:2014} confirmed the PWN nature of the system. The orientation of the nebula points back towards the centre of SNR MSH 11--61A, which suggests that it is the parent remnant of the pulsar. The whole system lies at an estimated distance $7 \pm 1$\kpc~\citep{Reynoso:2006} and has a possible age of approximately 20\kyr, which implies a projected pulsar kick velocity of about 1000\kms. The Lighthouse Nebula has the brightest, longest, and most distant X-ray jet known so far, with a length of about 10\pc~\citep{Dinsmore:2024}. The misaligned jets and the pulsar were not detected at radio wavelengths, but the tail of the nebula was clearly detected in observations at 2\,GHz with ATCA~\citep[Australia Telescope Compact Array;][]{Pavan:2014}.

\section{Data taking and processing}
\label{data}

\begin{figure}
\centering
\includegraphics[width=\columnwidth]{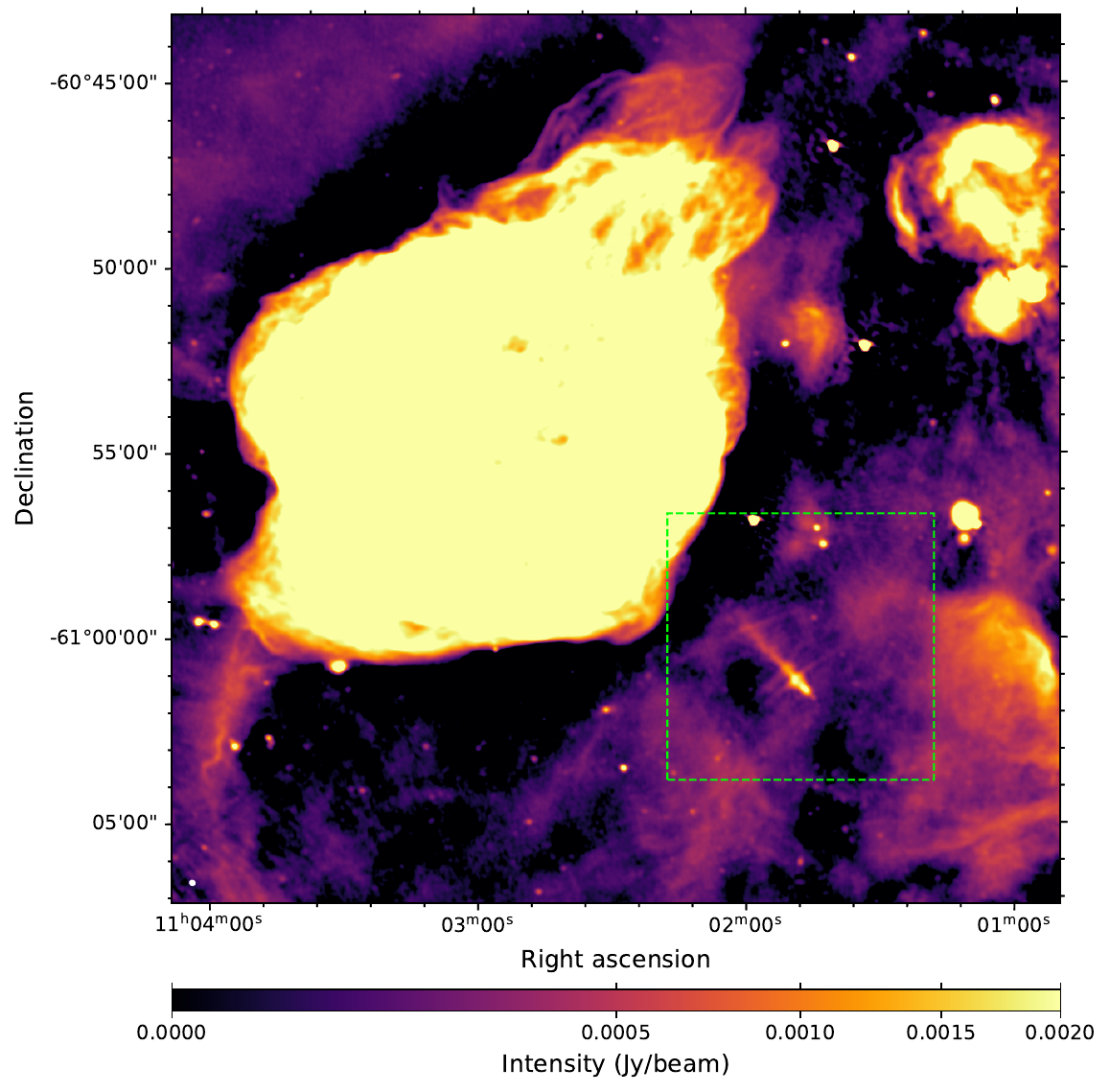}
\includegraphics[width=\columnwidth]{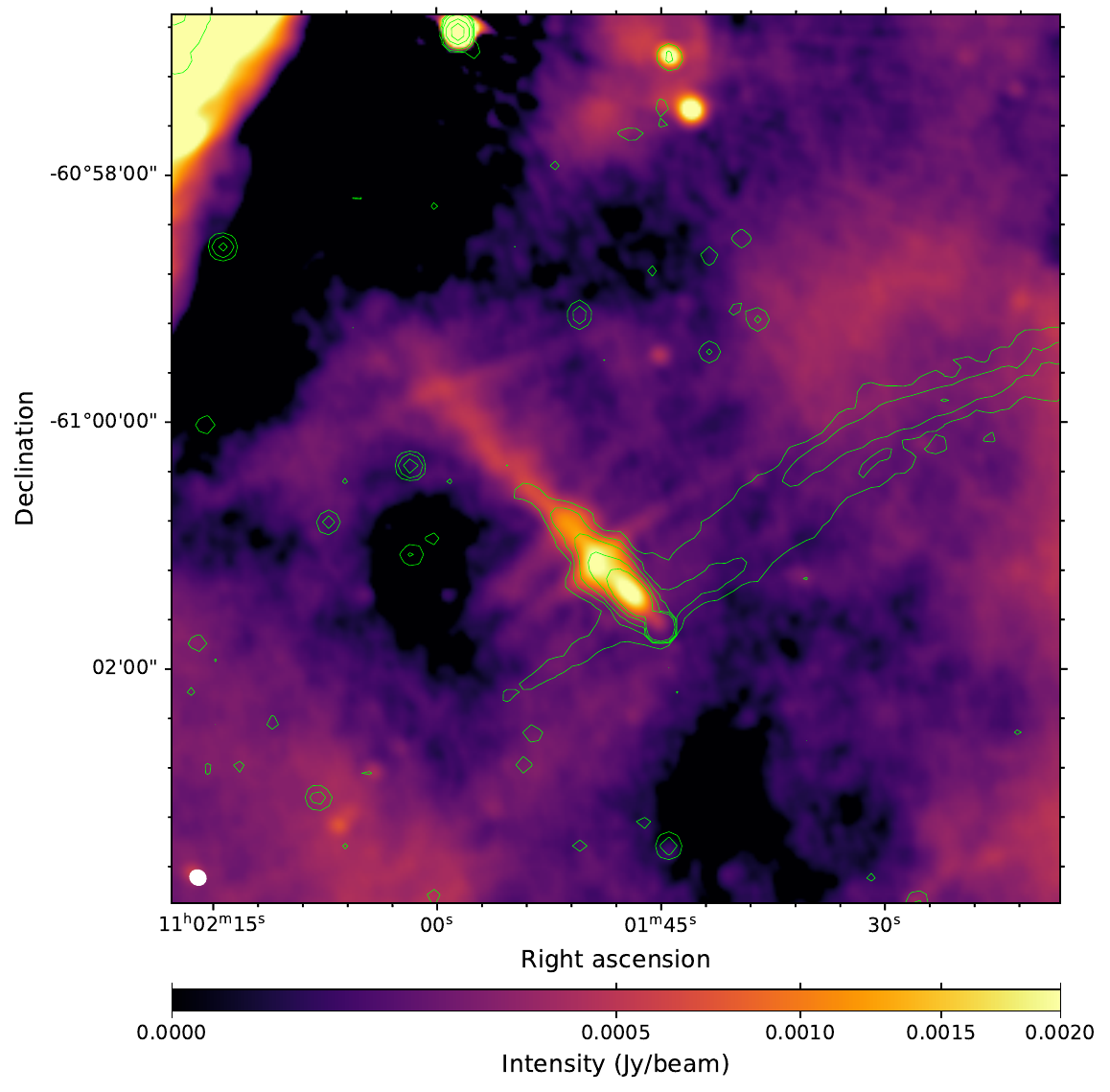}
\caption{Intensity map of the central part of the region covered by our MeerKAT 1.28\ghz observations. Top panel: Broad view dominated by the large, saturated source SNR~MSH~11--61A. Bottom panel: Zoomed-in view of the Lighthouse Nebula. The green contours correspond to the X-ray photon flux measured with Chandra/ACIS in the $0.5-7$\kev band. The beam size is indicated by a white ellipse in the bottom-left corner.}
\label{fig:intensity}
\end{figure}

\begin{figure}
\centering
\includegraphics[width=\columnwidth]{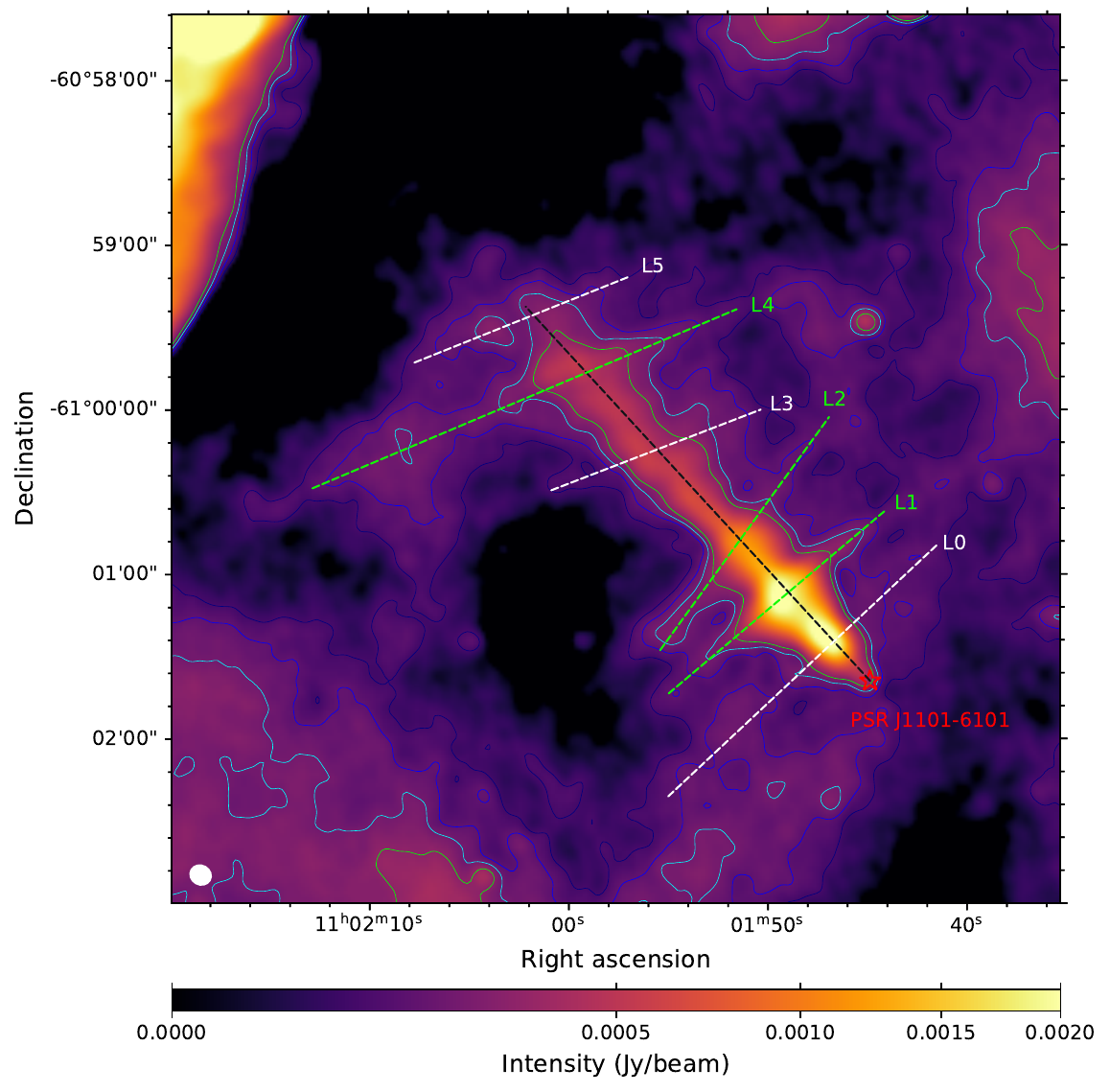}
\caption{Intensity map of a zoomed-in region centred on the Lighthouse Nebula, obtained from MeerKAT 1.28\ghz observations. The position of PSR J1101--6101 is marked by a red star, and the contours correspond to emission at significance levels of 3, 5, 7, and 10$\sigma$. The black dashed line indicates the longitudinal axis of the nebula, and the green and white dashed lines indicate conspicuous and tentative transverse emission streaks, respectively.}
\label{fig:chart}
\end{figure}

\subsection{MeerKAT}
\label{data:meerkat}

We observed the field of PSR~J1101$-$6101 with the MeerKAT radio telescope~\citep{Jonas:2016} in the L band (856--1712\,MHz) over two epochs. The first observation took place on March 16, 2021, with 63 antennas and a total integration time of 4.2\,hours, while the second was conducted on June 1, 2021, with 59 antennas for 3.7\,hours, yielding a combined on-source time of 6.3\,hours. Both observations used the 4k wide-band continuum mode, providing 4096 frequency channels across the band. We observed the primary calibrator J0408$-$6545 at the start and end of each session for flux density and bandpass calibration, while we observed the secondary calibrator J0906$-$6829 at regular intervals for time-dependent gain corrections. We reduced the data using standard procedures with the \textsc{CARACal} pipeline~\citep{Jozsa:2020}. We combined the calibrated measurement sets from both epochs and imaged them using \textsc{WSClean}~\citep{Offringa:2014} with Briggs weighting (robust~$=0$), applying multi-scale and multi-frequency synthesis deconvolution. The final image achieves a root-mean-square (rms) noise of $30$\ujy\,beam$^{-1}$ near the phase centre and a synthesised beam of $7.7\arcsec\times\,6.9\arcsec$ at a position angle of 60.5\deg.

\subsection{ASKAP}
\label{data:askap}

We also used ASKAP observations~\citep[]{2021PASA...38....9H} as from the Evolutionary Map of the Universe survey~\citep[EMU\footnote{\href{https://emu-survey.org/}{Evolutionary Map of the Universe~(EMU)}};][for review]{2025PASA...42...71H}. The EMU field covering the PSR~J1101--6101 region corresponds to scheduling block 60336 and was observed on March 25, 2024. The EMU data have a central frequency of 944\,MHz with a bandwidth of 288\,MHz, divided into 288 individual channels of 1\,MHz width. We convolved the images to a common angular resolution of 15$^{\prime\prime}\times$\,15$^{\prime\prime}$, with a typical rms sensitivity of about 30\,$\mu$Jy\,beam$^{-1}$.

We processed the observations using the standard ASKAPsoft pipeline~\citep{Guzman_Askapsoft},  following the procedures described by~\citet{2021PASA...38...46N}. We retrieved science-ready data products from the CSIRO ASKAP Science Data Archive~\cite[CASDA\footnote{\href{https://research.csiro.au/casda/}{CSIRO ASKAP Science Data Archive (CASDA)}};][]{2020ASPC..522..263H}. We used the ASKAP data to compute the spectral index map presented in Sect.~\ref{data:specmap} and Fig.~\ref{fig:specmap}.

\subsection{MeerKAT intensity maps}
\label{data:intensity}

\begin{figure}
\centering
\includegraphics[width=\columnwidth]{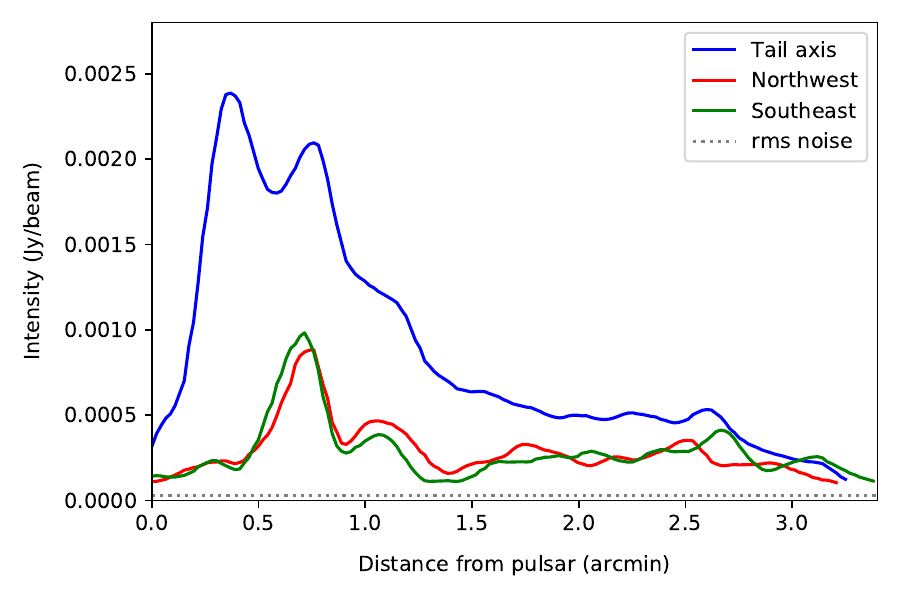}
\includegraphics[width=\columnwidth]{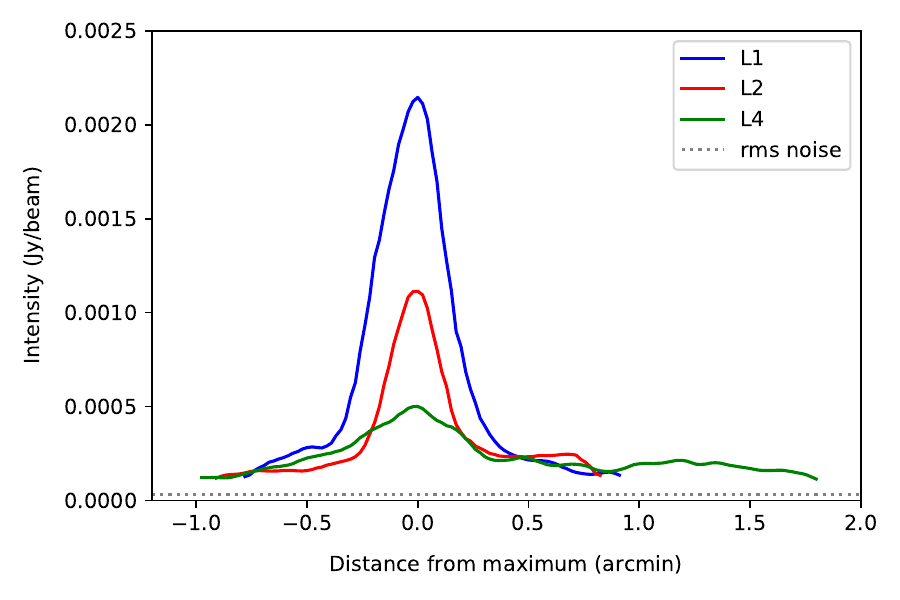}
\caption{Radio intensity profiles of the Lighthouse Nebula. Top panel: Intensity along the central axis of the tail and two directions shifted offset towards the north-west and south-east, respectively. Bottom panel: Intensity along the three main streaks.}
\label{fig:profiles}
\end{figure}

Fig.~\ref{fig:intensity} displays the radio intensity distribution in the Lighthouse Nebula region. The Lighthouse Nebula has an extended $\sim$6\pc radio tail that was partially detected in earlier ATCA observations~\citep{Pavan:2014}. In addition to providing a broader picture of that component, MeerKAT unveiled an array of streaks or filaments transverse to the tail. Three streaks are prominent and are marked in green in Fig.~\ref{fig:chart} (labelled L1, L2, and L4), while three others are more tentative and are marked in white (labelled L0, L3, and L5). Table~\ref{tab:streaks} lists the distances from the pulsar (defined as the intersection of the black and green or white lines in Fig.~\ref{fig:chart}) and the length of each streak (decomposed into its northern and southern parts).

The top panel of Fig.~\ref{fig:profiles} displays intensity profiles along three lines along the tail: one starting at the pulsar and extending to the end of the detectable tail, and two parallel lines shifted by 12\arcmin\ in the northwest and southeast that probe the intensity on either side of the tail, where the streaks connect. These profiles highlight clear intensity enhancements at approximately 0.7\arcmin\, 1.1\arcmin\, and 2.7\arcmin\ from the pulsar, corresponding to L1, L2, and L4, respectively. The tentative streaks L0, L3, and L5 appear as more modest signal increases at 0.3\arcmin\, 1.7\arcmin,  and 3.1\arcmin.

The bottom panel of Fig.~\ref{fig:profiles} displays intensity profiles along the main streaks L1, L2, and L4. We rescaled the positions so that the peak intensity is located at zero, with positive distances corresponding to the southeast direction. The plot illustrates that the intensities are very similar in all three streaks on either side of the tail, despite their different distances from the pulsar.

\begin{table}[!t]
\centering
\caption{Angular position and extent of the radio streaks.}
\begin{tabular}{| c | c | c |}
\hline
  & Distance from pulsar & Length (North+South) \\
  & (\arcsec) & (\arcsec) \\
\hline
L0 & 20 & ~51 + 84 \\
L1 & 44 & ~45 + 59 \\
L2 & 70 & ~54 + 51 \\
L3 & 114 & ~39 + 42 \\
L4 & 152 & 108 + 58 \\
L5 & 181 & ~35 + 49 \\
\hline
\end{tabular}
\label{tab:streaks}
\end{table}

\begin{figure}
   \centering
   \includegraphics[width=0.9\columnwidth]{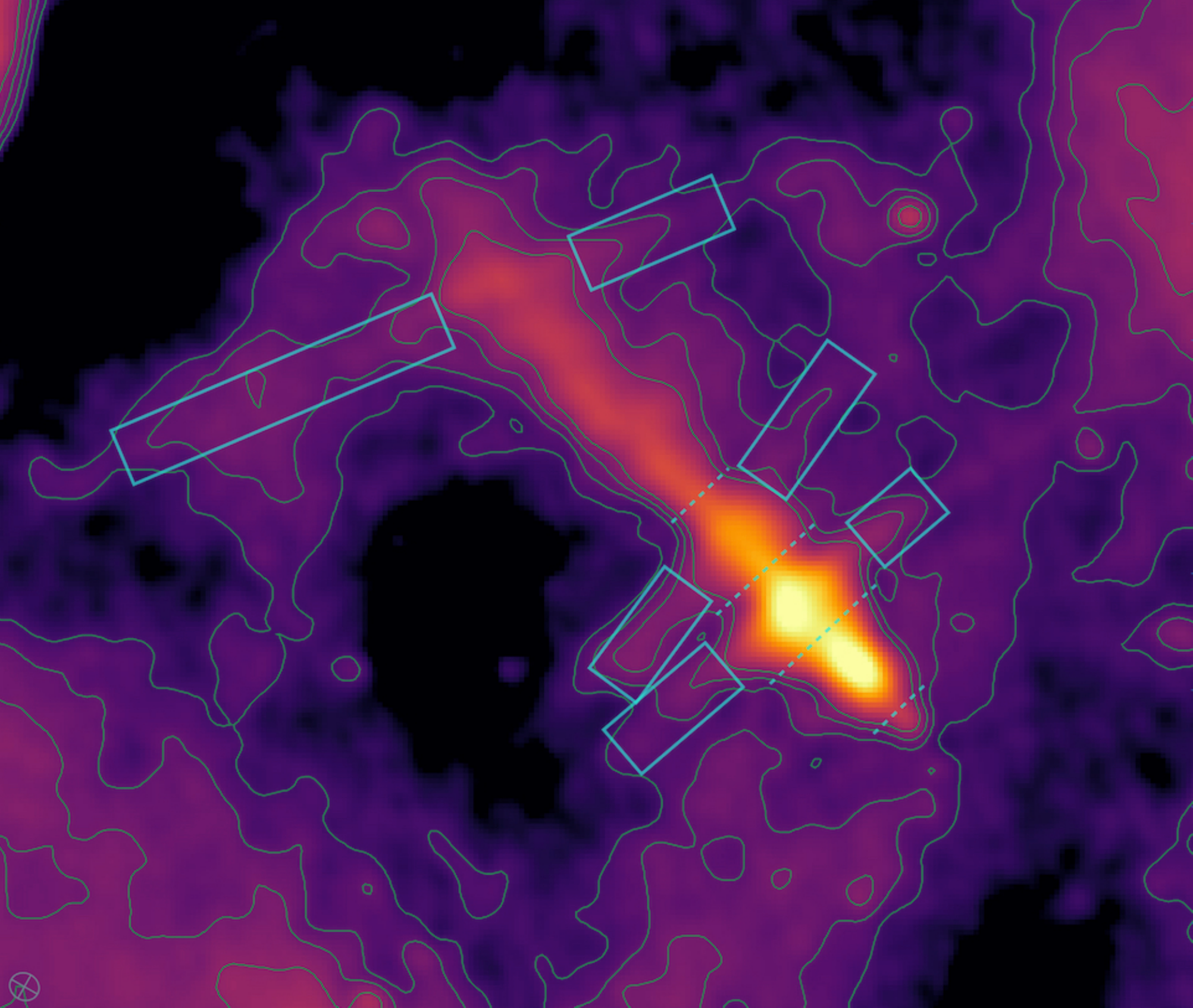}
   \caption{Regions used for flux extraction. The blue boxes correspond to the northern and southern parts of the main streaks L1, L2, and L4. The blue dashed lines divide the tail into five regions: pulsar, head, shoulder, hip, and leg~(from bottom-right to top-left).}
   \label{fig:regions}
    \end{figure}

 Table~\ref{tab:flux} lists the fluxes from selected regions of the source, together with their uncertainties~(see Fig.~\ref{fig:regions} for the locations of the different regions). We estimated the uncertainties as the average standard deviation of the residual signal over the region footprint within adjacent source-free areas. For the longitudinal tail, we extracted the fluxes within the 10$\sigma$ contour for four regions: one for each emission maximum downstream of the pulsar and one for the remainder of the tail. In addition, we extracted the flux density of the weak emission feature coincident with the pulsar. For the main transverse streaks, we extracted the flux in rectangular regions 15\arcmin\ wide, starting at the 10$\sigma$ contour and approximately encompassing the $>3\sigma$ flux. Table~\ref{tab:flux} gives the values as the sum of two terms corresponding to the flux on the southeastern and northwestern sides of the longitudinal tail, respectively. The total flux in each streak is in the range $\sim2-6$\mjy, which implies that the total flux in all streaks is comparable to that of the brightest parts of the tail, such as the head or shoulder.  

\begin{table}[!t]
\centering
\caption{Flux densities for selected regions in the nebula (see Fig.~\ref{fig:regions}).}
\begin{tabular}{| c | c |}
\hline
Region & Flux density~(mJy) \\
\hline
Pulsar   & $0.42 \pm 0.05$ \\
Head     & $10.8 \pm 0.4$  \\
Shoulder & $14.3 \pm 0.5$  \\
Hip      & $8.1  \pm 0.4$  \\
Leg      & $14.8 \pm 1.1$  \\
\hline
L1-SE + L1-NW & $(1.3 \pm 0.3)+(1.0 \pm 0.2)$ \\
L2-SE + L2-NW & $(1.6 \pm 0.3)+(1.5 \pm 0.3)$ \\
L4-SE + L4-NW & $(3.9 \pm 0.7)+(1.6 \pm 0.3)$ \\
\hline
\end{tabular}
\label{tab:flux}
\end{table}

\subsection{ASKAP and MeerKAT spectral-index map}
\label{data:specmap}

To derive the spectral index distribution across the Lighthouse Nebula and its associated uncertainty, we used ASKAP--EMU 0.94\ghz and MeerKAT 1.28\ghz continuum images, following the procedure of \cite{2026arXiv260622354L}. The spectral index was computed pixel-by-pixel. We first regridded the images to the finer MeerKAT pixel scale~(1.3$^{\prime\prime}\times$\,1.3$^{\prime\prime}$), ensuring that each pixel sampled the same sky area at both frequencies,  and then convolved them to a common angular resolution of 15$^{\prime\prime}\times$\,15$^{\prime\prime}$ to match the larger ASKAP--EMU beam.

\begin{figure}
\centering
\includegraphics[width=0.9\columnwidth]{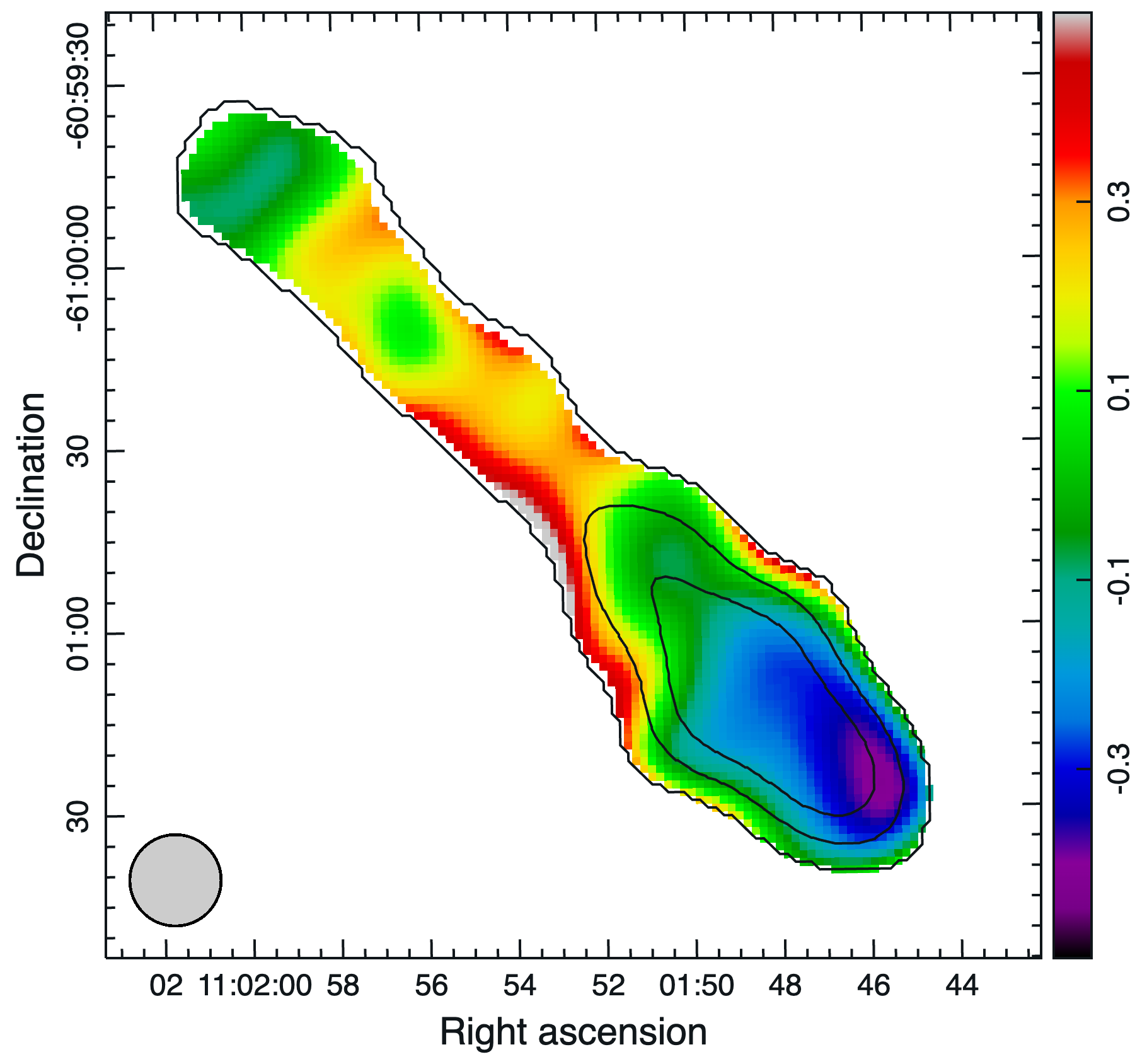}
\includegraphics[width=0.9\columnwidth]{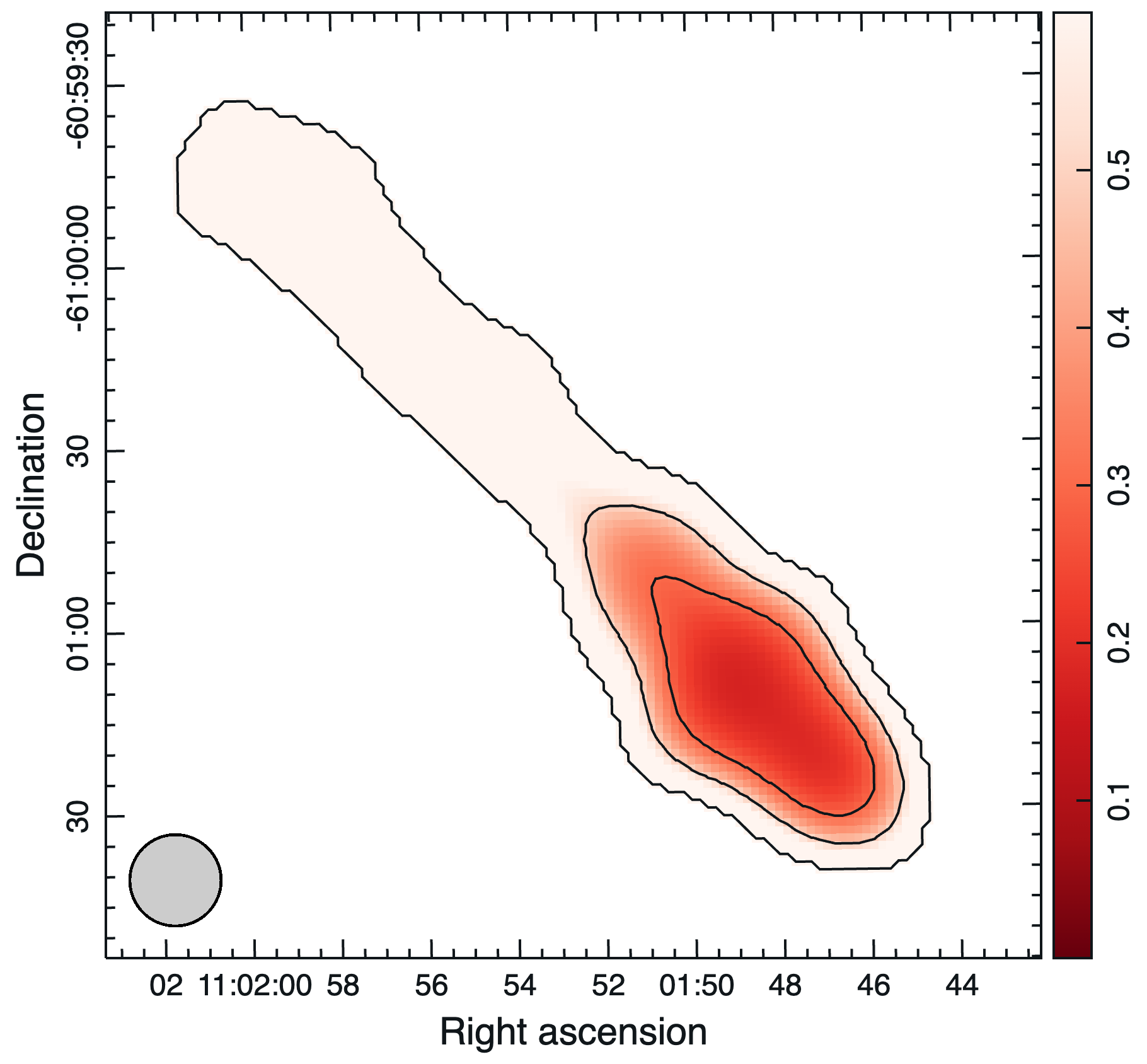}
\caption{Spectral index map of the Lighthouse Nebula over the 0.94\ghz and 1.28\ghz range~(top panel) and the corresponding statistical uncertainty, $\sigma_{\alpha, stat}$~(bottom panel). Black contours overlaid on both maps indicate $\sigma_{\alpha, stat}$ = 0.3, 0.5, and 1. The synthesised beam of 15$^{\prime\prime}\times$\,15$^{\prime\prime}$ is indicated by the black circle in the lower-left corner.}
\label{fig:specmap}
\end{figure}

Given that only two frequencies are available, we derived a two-point spectral index, defined as $\alpha$ = $[\ln(\nu_1 / \nu_2)]^{-1}$ ${\rm ln}(S_2/S_1)$, where $S_1$ and $S_2$ correspond to the flux densities at $\nu_1$ = 0.94\ghz and $\nu_2$ = 1.28\ghz. To ensure reliable measurements, we excluded pixels with flux densities below the 5$\sigma$ threshold in either band from the analysis. We estimated the statistical uncertainty in the spectral index using standard error propagation, $\sigma_{\alpha, stat}$ = $|\ln(\nu_1 / \nu_2)|^{-1}$$\sqrt{(\sigma_1/S_1)^2 + (\sigma_2/S_2)^2}$, where $\sigma_i$ is the rms noise of each image. The relative flux-density calibration of the two instruments introduces an additional systematic uncertainty. We adopted conservative flux-scale uncertainties of 10\% for ASKAP and 5\% for MeerKAT \citep{2024MNRAS.531..649G}, which give $\sigma_{\alpha, sys}$ $\simeq$ 0.36. We did not include this systematic uncertainty in the pixel-by-pixel uncertainty map, as it primarily represents an overall offset in the spectral-index scale and affects the absolute values of $\alpha$ more strongly than their relative spatial variations across the nebula.
Fig.~\ref{fig:specmap} shows the resulting spectral index and statistical uncertainty maps, with contours at $\sigma_{\alpha, stat}$ = 0.3, 0.5, and 1 overlaid on both maps. We masked pixels with $\sigma_{\alpha, stat} >$ 1 and set them to NaN.

\begin{figure*}
\includegraphics[width=\textwidth]{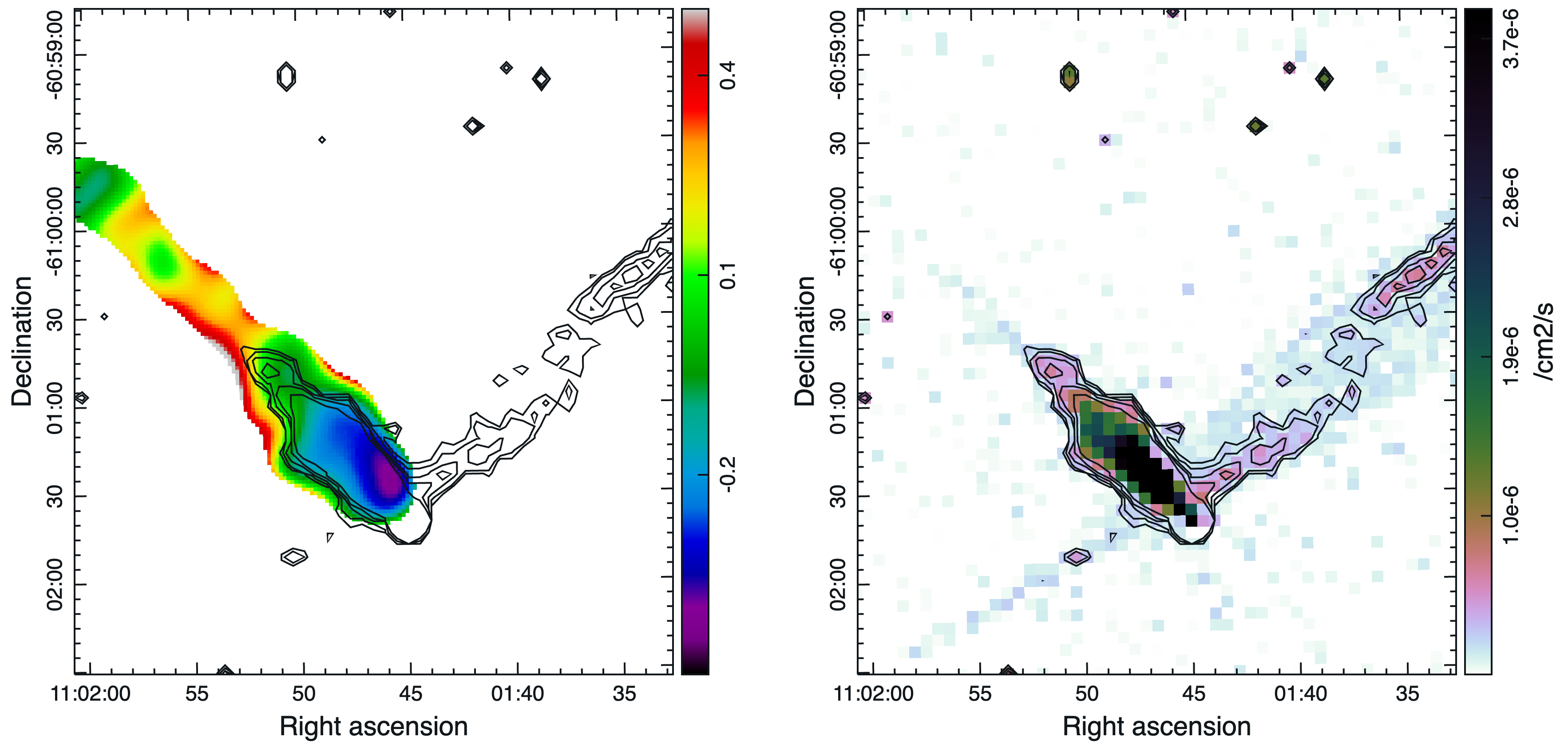}
\caption{Spectral index map~(left panel) and Chandra X-ray flux image~(right panel). Contours correspond to the X-ray image.}
\label{fig:specmap_contours}
\end{figure*}

\subsection{Summary of observables}
\label{data:summary}

The processed ASKAP and MeerKAT data, including both intensity and spectral index maps, reveal several observational properties. We broadly divide the radio emission  into three components: a bright section extending to $\sim$$2.5$\pc from the pulsar, a fainter downstream section, and a set of transverse streaks. 

\paragraph{Near-pulsar tail region}
The radio emission immediately trailing the pulsar exhibits the following properties:
\begin{enumerate}
    \item A faint emission component at the pulsar position that extends $\sim$0.5\pc downstream towards the north-east~(Fig.~\ref{fig:regions}).
	\item A strong rise in radio intensity leads to a first maximum (the head), followed by a dip and a second maximum (the shoulder). The emission then decreases rapidly before reaching a third maximum and plateau (the hip). The separation between these maxima is approximately 1\pc, comparable to the lateral extent of the tail.
	\item The spectral index map exhibits an apparent asymmetric distribution, with the steepest values located on the side slightly downstream of the misaligned X-ray jet~(Figs.~\ref{fig:specmap} and \ref{fig:specmap_contours}). Given the statistical and systematic uncertainties discussed in Sect.~\ref{data:specmap}, the absolute spectral-index values should be interpreted with caution.
\end{enumerate}

The first maximum in radio emission coincides with the peak of the X-ray $0.5-10$\,keV band emission~\citep[][see also the bottom panel of Fig.~\ref{fig:intensity}]{Pavan:2014}.
The spectral-index values in this region lie between --0.3 and 0.0, within the range typically observed in BSPWNe~\citep{Kothes:2017}. The origin of the spectral asymmetry and steepening near the base of the X-ray jet remains unclear and requires further investigation, as it may provide constraints on the mechanisms responsible for the formation of the X-ray jet.

\paragraph{Downstream tail region}
At larger distances from the pulsar, the dimmer part of the tail (the leg) is characterised by:
\begin{enumerate}
    \item a progressive decrease in intensity over 3.5\pc~ (Fig.~\ref{fig:regions}).
	\item a noticeable wobbling of the tail around its longitudinal axis~ (Fig.~\ref{fig:regions}).
    \item a flattening or inversion of the spectral index, with nominal values between 0.0 and 0.4, although the uncertainties are large in this faint region~(Fig.~\ref{fig:specmap}).
\end{enumerate} 

The observed wobbling around the longitudinal axis is reminiscent of the zig-zag behaviour seen in the PWN inside SNR\,N206 in the Large Magellanic Cloud~\citep{2025SerAJ.211...27G}. A similar effect was also observed in ATCA 5\ghz images of the Potoroo nebula~(S. ~Lazarevi\'c, private communication).

The spectral hardening at larger distances from the pulsar is somewhat intriguing~(Fig.~\ref{fig:specmap}), as it is opposite to the behaviour observed in Potoroo~\citep{Lazarevic:2024} and to what is expected from particle radiative cooling~(although gigahertz-emitting particles are not expected to cool significantly over such scales; see Sect.~\ref{interp:relic}). One possible explanation is the progressive mixing of the shocked pulsar wind with the ambient material due to disruption of the tail, concomitant with the observed streak-like structures. Alternatively, the derived spectral index might be affected by the different sensitivities of the ASKAP and MeerKAT observations~(survey versus dedicated observations), particularly in the faintest part of the source. This possibility is supported by the relatively large uncertainties in this region~(Fig.~\ref{fig:specmap}, bottom panel).

\paragraph{Transverse streaks}
The newly detected radio streaks or filaments transverse to the tail~(Fig.~\ref{fig:chart}) have the following properties:
\begin{enumerate}
	\item Some streaks coincide with brightening regions along the tail~(L0, L1, and L2), while one corresponds to a lateral broadening~(L1).
	\item Streaks appear on both sides with similar lengths and show apparent continuity from one side of the tail to the other.
	\item Streaks do not correspond to the known X-ray jet or counter-jet system, and none originate from the bow-shock apex.
	\item The separation between streaks along the tail~($\sim$1\pc) is comparable to the lateral extent of the tail.
	\item No significant evolution in their intensity, length, or width is observed along the pulsar trajectory.
\end{enumerate}

The spatial coincidence between the lateral streaks and the radio-emission maxima along the tail suggests that these faint features are not spurious artefacts introduced during data reduction.

\section{Interpretation}
\label{interp}

\subsection{Relation of the radio streaks to the X-ray jets}
\label{interp:relic}

The radio structures could represent a relic trace of the same feature observed in X-rays close to the pulsar.
However, this scenario can be readily ruled out. 
If the radio emission from the streaks originates from the same particles that once emitted in X-rays, the particles must have lost enough energy during the time required for the pulsar to move from a streak to its present position to emit photons in the radio band.
In the $\sim$$30\,\mu$G amplified magnetic field~($\delta B$) of the X-ray feature~\citep{Pavan:2014,Olmi:2024}, this requires the particle energy to decrease from $40-100$~TeV to $3-4$ GeV. However, this requires a much longer time than the travel time between any of the streaks and the current pulsar position.
Considering the first streak, L1, this travel time is $t_{\mathrm{X-L1}}\simeq 1780-590$ yr, where the interval is set by the range of possible pulsar speeds~($\vpsr=800-2400$ km\,s$^{-1}$).
The most favourable viable scenario for interpreting the radio streaks as relics of the X-ray features is one in  which the amplified magnetic field $\delta B_X$ survives for $t_{\mathrm{X-L1}}$ before eventually
dropping to the ISM value $B_{\mathrm{ISM}}=3\,\mu$G.
The $40-100$ TeV particles, originally emitting X-ray photons in $\delta B_X$, will then produce UV light, to which ISM absorption makes us largely insensitive, both in the amplified field and in the ambient one.
In fact, one expects the amplified magnetic field to decay as soon as the pulsar moves away and particle injection in the flux tube stops. In this case, the synchrotron loss time becomes even longer, and particles that originally emitted X-rays would still emit X-rays, but at a level difficult to detect over the background noise~(signal-to-noise $\sim 3$ at best).

A related possibility is that the original X-ray feature already contains radio-emitting particles and that it becomes a radio feature once the magnetic field decays sufficiently to prevent any X-ray emission.
On the one hand, it is difficult to explain how to extract low-energy particles efficiently from the bow-shock apex to populate the filament, given their much smaller Larmor radii~($10^{-4}-10^{-5}$) compared to those of the X-ray emitting particles~\citep{Olmi:2019}. 
On the other hand, even in this scenario, the X-ray feature should produce detectable radio emission. We quantify this statement in the following section.
It then becomes apparent that the nature of the particles responsible for these two types of structure~(X-ray filaments and radio streaks) is different.

\subsection{Synchrotron-emitting particle populations in the system}
\label{interp:particles}

A relevant piece of information comes from the measured spectral indices~(radio and X-rays) in the PWN~(see the previous Sect.~\ref{data:specmap} and~\citealt{Pavan:2016}) and in the X-ray feature~\citep{Pavan:2016, klingler:2023}.

In general, the emission spectrum of a PWN is well explained by assuming the injection of two different particle populations, both described as power-law spectral distributions, but with different spectral indices below and above a Lorentz factor $\gamma_B$, namely, $f(\gamma,t)=N_0(t)(\gamma/\gamma_B)^{-p_i}$, where $p_i\equiv p_L$ for $\gamma\leq \gamma_B$, while $p_i\equiv p_H$ for $\gamma > \gamma_B$.
The spectral index of synchrotron radiation $\alpha$~(now defined as $S_\nu \propto \nu^{-\alpha}$) is related to that of the emitting-particles distribution $q$ as $\alpha=(q-1)/2$; $q$ is in turn related to the injection spectral index as $q=p$ at low enough energy, where synchrotron losses are not severe~($\gamma<\gamma_s$),
and $q=p+1$ for particles that radiate effectively~($\gamma>\gamma_s$), with $\gamma_s$ being the Lorentz factor corresponding to the synchrotron cooling break.
Spectral analysis of several PWNe indicates $\gamma_B\sim10^5-10^6$~\citep[e.g. ][]{Bucciantini:2011,Torres:2014} and $\alpha_{\mathrm{low}}\sim0-0.3$, while $\alpha_{\mathrm{high}}\sim 0.6$.

It is notable that, in the case of the Lighthouse PWN, the measured photon index~($\Gamma=\alpha+1$) is $\Gamma_X\simeq 1.7-2.5$~\citep[see][their Fig.~6]{Pavan:2016} up to a distance of $\sim50^{\prime\prime}$ from the pulsar, which is approximately where the first bright part of the tail~ (between L1 and L2) ends.
The photon index then softens with increasing distance from the pulsar, suggesting synchrotron cooling along the main tail. The average value in that region is $\Gamma_X\sim 2.1$ in terms of the photon index, implying a spectral index $\alpha_X\sim 1.1$, compatible with a cooled spectrum.

However, the radio spectral index extracted from the PWN varies between -0.4 and +0.4, but the spectrum could be much steeper further downstream, as observed in other similar sources~\citep[e.g.][]{Lazarevic:2024}.
This suggests two possible interpretations of the observed spectrum: i) a standard description with two particle populations, including a flat low-energy component~($\alpha_L=0.0$); and ii) a spectrum described by a single population with $\alpha\simeq 0.6$. 
In this second case, if there is a low-energy flat-spectrum population, the intrinsic break must be below a few $\times 10^3$~(corresponding to $\sim$ GeV), so that this component does not appear in the emission spectrum.
In both cases, a synchrotron break~($\gamma_s$) must occur between the MeerKAT and Chandra observational bands to allow for a 1/2 steepening of the spectral index due to radiation cooling, leading to a cooled value of $\alpha\simeq 1.1$.

The situation is different for the X-ray feature, where the measured photon index is compatible with the uncooled value: $\Gamma_X\sim1.3-1.8$ in the first $150^{\prime\prime}$ from the pulsar. 
This matches the expectation that the particles populating the X-ray feature escape the BSPWN before starting to cool in the PWN and that in the feature, we observe their injection spectrum~\citep{Olmi:2019}.

\begin{figure}
\centering
\includegraphics[width=\columnwidth]{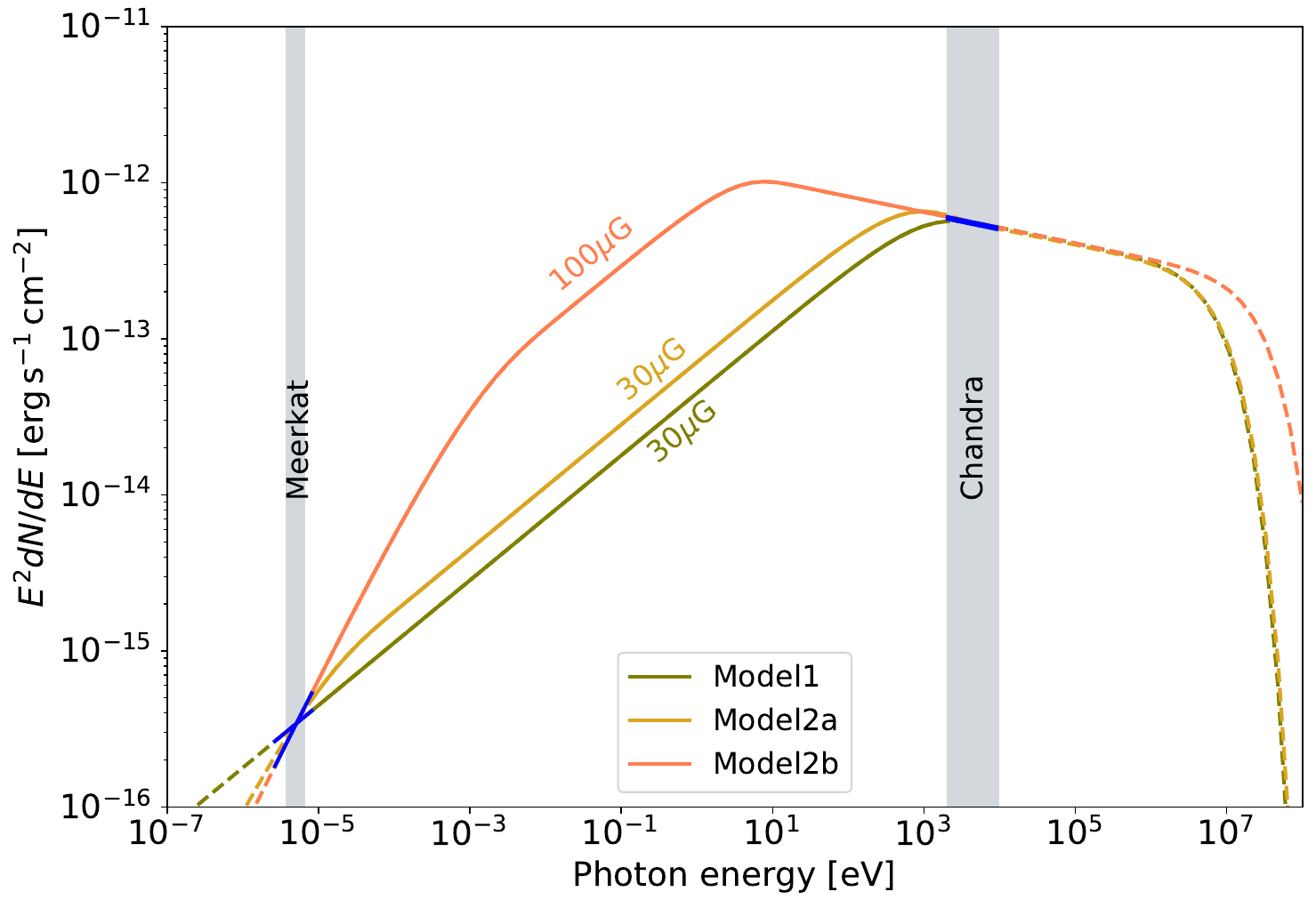}
\caption{Spectral energy distribution~(SED) computed using different models for the particle injection spectrum. Model1 assumes a single particle population~and a radiation spectral index of $\alpha=0.6$, while Model2a and Model2b assume two populations~with the same radiation spectral indices $\alpha_L=0.0$ and $\alpha_H=0.6$ but different positions of the intrinsic and synchrotron breaks~($\gamma_b$ and $\gamma_s$) depending on the nebular magnetic field strength. The latter is reported on the plot using the same colour as the corresponding curve. Observational bands are shown as grey-shaded regions. Blue segments indicate the spectral slopes in the observed windows.}
\label{fig:sed}
\end{figure}

We used the particle spectral distributions discussed above to produce different possible PWN spectra\footnote{We produced synchrotron spectral models using the Naima Python library~\citep{zabalza:2015}.} to fit the available radio and X-ray data~(from MeerKAT and Chandra).
We fixed the observation energies and detected fluxes and took into account the large range of possible pulsar velocities~(800--2400~km\,s$^{-1}$) inferred from the association with the nearby young SNR~\citep{garcia:2012}. We then determined the magnetic field in the PWN, the injection break $\gamma_B$, and the synchrotron break $\gamma_s$ for each scenario. 
We summarise our results in Fig.~\ref{fig:sed} and Table~\ref{tab:sp_models}~(Model1 for a single particle population and Model2 for two populations at injection).

\begin{table}
\caption{Summary of the spectral modelling parameters.}
\begin{center}
\fontsize{9}{9}\selectfont
\begin{tabular}{c l c c c}
\hline\hline
 &  & Model1 & \multicolumn{2}{c}{Model2} \\
\cline{4-5}
 &  &  & 2a & 2b \\
\hline
\multirow{10}{*}{\rotatebox{90}{$\vpsr=2400$ km/s}} 
 & $\alpha_L$ & --   & 0.0 & 0.0 \\
 & $\alpha$   & 0.6 & 0.6 & 0.6 \\
 & $\gamma_s$ & $5.4\times10^7$ & $5.4\times10^7$ &  $4.0\times10^6$ \\
 & $\nu_s$ [Hz] & $9.7\times10^{16}$ & $9.7\times10^{16}$ & $2.0\times10^{15}$ \\
 & $\gamma_B$ &  -- & $6.3\times10^3$ & $3\times10^4$  \\
 & $\nu_B$ [Hz] & -- & $1.3\times10^9$ &  $1.1\times10^{11}$ \\
 & $B$ [$\mu$G] & 30 & 30 & 100 \\
 & $E_{\mathrm{part}}$ [erg] & $6.5\times10^{45}$ & $1.3\times10^{45}$ & $2.7\times10^{45}$ \\
 & $N$ [particles] & $2\times 10^{48}$ & $4\times 10^{47}$ & $1.4\times 10^{47}$  \\
 & $\kappa$ & $2\times10^5$ & $5\times10^4$ & $2\times10^4$ \\
\hline
\multirow{10}{*}{\rotatebox{90}{$\vpsr=800$ km/s}} 
 & $\alpha_L$ & --   & 0.0 & 0.0 \\
 & $\alpha$   & 0.6 & 0.6 & 0.6 \\
 & $\gamma_s$ & $7.7\times10^7$ & $7.7\times10^7$ &  $5.8\times10^6$ \\
 & $\nu_s$ [Hz] & $9.7\times10^{16}$ & $9.7\times10^{16}$ & $2.0\times10^{15}$ \\
 & $\gamma_B$ &  -- & $6.3\times10^3$ & $3\times10^4$  \\
 & $\nu_B$ [Hz] & -- & $6.5\times10^8$ &  $5.3\times10^{10}$ \\
 & $B$ [$\mu$G] & 13 & 13 & 49 \\
 & $E_{\mathrm{part}}$ [erg] & $2.0\times10^{46}$ & $1.3\times10^{45}$ & $7.6\times10^{45}$ \\
 & $N$ [particles] & $6.5\times 10^{48}$ & $9.8\times 10^{47}$ & $2.8\times 10^{47}$  \\
 & $\kappa$ & $7\times10^5$ & $10^5$ & $3\times10^5$\\
\hline
\end{tabular}
\end{center}
\label{tab:sp_models}
Notes: Model1 assumes a single power-law particle population with spectral index $\alpha$, while Model2a and Model2b assume a broken-power-law distribution, with the addition of the low-energy spectral index $\alpha_L$ and varying positions of the break. We denote the total particle energy within the PWN by $E_{\mathrm{part}}$. The cases shown in Fig.\ref{fig:sed} correspond to $\vpsr=2400$ km/s.
\end{table}
We limited the possible models to cases consistent with a magnetic field below $\sim$100\,$\mu$G.
Additionally, we find that no fit to the data is possible assuming $\alpha>0.6$.

This is a simplified description based on one-zone modelling, which captures only the average magnetic field required to reproduce the observed fluxes. The magnetic field strengths considered are close to the equipartition value $\sim$40\,$\mu$G, which we estimated from radio and X-ray emission using the expression by~\cite{Pacholczyk:1970}.
An important assessment of the reliability of this picture requires computing:
i) the fraction of energy that goes into illuminating the system compared with the available energy injected by the pulsar; and ii) the number of particles needed to feed the PWN and the features. 

The energy injected by the pulsar in each region of the system within a distance $\Delta d$ from the pulsar can be roughly estimated as $E_{\mathrm{inj}}\sim \dot{E\,}\Delta d/V_{\mathrm{PWN}}$. For example, the energy injected into the main~(brighter) part of the PWN within a distance $\Delta d \simeq d_1$, corresponding to the distance between the pulsar and L1$\simeq 1.5$ pc, is $E_{\mathrm{inj,1}}\sim 4\times 10^{46}$ erg. 
The energy needed to power each region of the system can be obtained by integration of the magnetic-field and particle-energy densities derived from fitting the emission.
We find that a fraction of $\sim 20\%$ of the available energy goes into accelerated particles, a value compatible with what is expected for PWNe; for the Crab Nebula, for example, ~\citep{Hester:2008} estimated a fraction of $\sim 30\%$.
Powering the entire system requires a fraction of $\sim 40\%$, considering the main PWN region up to distance $d_1$, the first streak, and the X-ray feature.

We computed the number of particles injected into each part of the system~(PWN, feature, and streaks), $N$, from the normalisation of the injection spectrum, which we determined by comparing the modelled and observed fluxes in the different regions. The resulting number of particles depends on the spectral model and assumed $\vpsr$, ranging from $N\sim 2\times 10^{47} $ to $ 7\times 10^{48}$ particles.
From this, we obtain the particle injection rate, $\dot{N}=N\, V_{\mathrm{PSR}}/\Delta d$.
We then compared this with the theoretical pair production rate $\dot{N}_{\mathrm{GJ}}=(2\pi /P)^2 B_* R_*^3/(ec)$~\citep{GJ:1969}, where $P$ is the pulsar period, 
$B_*$ is the magnetic field at the star surface, 
and $R_*\sim 10$ km is the pulsar radius, to estimate the pulsar pair multiplicity ($\kappa$) required to support our interpretation. 
We obtain $\kappa\simeq \dot{N}/\dot{N}_{\mathrm{GJ}}\sim 10^4-10^6$, with the extremes of the interval corresponding to the most extreme cases under consideration: $\vpsr=800$ km/s, and an injection spectrum with two populations implying an ambient field of $\sim 100\,\mu$G; or $\vpsr=2400$ km/s, and an injection spectrum with a single population.
These estimates are consistent with
theoretical expectations~\citep[see e.g. the discussion in][and references therein]{Amato:2014} and are close to the average value inferred from the modelling of well-studied PWNe \cite{Bucciantini:2011, Torres:2014}.

\subsection{Possible physical origin for the radio streaks}
\label{interp:origin}

In our interpretation, the most realistic origin of the radio streaks is that of dynamically induced features.
The tail of a BSPWN is expected to become dynamically unstable at a certain distance from the pulsar~\citep{Olmi:2019a}. Moreover, the shape of the main PWN region (up to a distance of $\sim d_1$) resembles that observed in other systems, such as the Guitar Nebula~\citep{Cordes:1993}.
The widening of the tail at the shoulders can be interpreted as the effect of mass loading by neutral atoms from the ambient 
\citep{Morlino:2015}. This not only modifies the tail shape, but also favours the formation of instabilities~\citep{Olmi:2018}.
It is impossible to constrain the mass loading in the Lighthouse Nebula, since only an upper limit exists on the emission of H$\alpha$, 
$F_{H\alpha}\lesssim 1.3 \times10^{-17}$ erg/s/cm$^2$/arcsec$^2$~\citep{Pavan:2016}, which is compatible with any neutral fraction.

Once instabilities disrupt the tail structure, a large number of particles can be injected into the ambient medium.
In this proposed dynamically driven  scenario, all particles in the tail are expected to escape with the same probability, unlike those in the X-ray feature.
We must then account for the non-detection of the streaks in X-rays. However, the expected X-ray flux from each streak, computed using the spectral model described in Sect.~\ref{interp:particles}, is comparable to the background noise in the Chandra data. 
Moreover, this scenario does not predict charge separation among the escaping particles, unlike the reconnection-driven escape at the base of the X-ray feature. This distinction is important because the radio streaks appear double-sided, with both sides having almost the same extension, unlike the X-ray features, which display a strong asymmetry associated with the charge-separated escape~\citep{Olmi:2019}.

Once they escape, the particles interact with the ambient magnetic field. The longitudinal extension ($l_i$) of the streaks cannot be interpreted as an effect of synchrotron cooling: the magnetic field required for such an interpretation is of order tens of milligauss, which is not realistic. 
If particle propagation is governed by diffusion, then $l_i$ relates to the local diffusion coefficient $D$ as $l_i\simeq \sqrt{D\Delta t_i}$, where the characteristic time involved can be assumed to be the time required for the pulsar to travel a distance $d_i$ ($\Delta t_i=d_i/\vpsr$).
From the previous relation we estimate $D\simeq 2-6\times 10^{27}$ cm$^2\,$s$^{-1}$, for $\Delta t_1=d_1/\vpsr$; this is an order of magnitude lower than the standard Galactic diffusion coefficient at the energy of the emitting particles, and it implies a magnetic-field perturbation of  $\delta B/B\simeq \sqrt{c R_L(E_\mathrm{MeerKAT})/(3D)}\sim 10^{-3}-10^{-2}$ on a scale $R_L(E_\mathrm{MeerKAT})$, corresponding to the Larmor radius of the particles emitting in the MeerKAT energy band.
From studies of cosmic-ray diffusion in the Galaxy~\citep{evoli:2021}, we estimate a diffusion coefficient on the scale of interest,  $R_L(E_\mathrm{MeerKAT})$ of $D_{\mathrm{Gal}}= 7.9\times 10^{28}$ cm$^2\,$s$^{-1}$, corresponding to an expected turbulence level of  $\delta B/B\simeq 10^{-3}$, which is compatible with the required value.
It is therefore plausible that normal ISM turbulence alone is responsible for the local modification of the magnetic field required to explain the extension of the radio features.

Given the relatively small value inferred for the diffusion coefficient in the streaks, another possibility is that the escaping particles produce their own turbulence via a self-generated instability, in particular, the resonant streaming instability. Since the radio features are thought to be produced by escaping particles that are not charge-separated, there is no net current to excite the non-resonant instability, as theorised for the X-ray feature (see~\cite{OrusaSironi26} for a related discussion).
Interestingly, if we interpret the length of the streaks as the distance travelled by particles drifting at speed $v_d$ in a time $\Delta t_i$, $v_d$ turns out to be super-Alfvènic ($v_d\sim 2000$ km/s) for any reasonable estimate of the local Alfvèn speed ($v_A\lesssim 200$ km/s), suggesting that the growth of the resonant streaming instability is likely limited by damping processes. The data currently at our disposal do not permit us to investigate this scenario in detail, but a better characterisation of the newly detected streaks from future, deeper radio observations may allow us to go further. More accurate intensity profiles and spectral-index measurements along the streaks could be compared with estimates of the growth and damping rates of the streaming instability, thereby providing an interesting test bed for fundamental processes of cosmic-ray transport~\citep[similar to synchrotron harps at the Galactic centre;][]{Thomas:2020}.

\section{Summary and conclusion}

In this work, we present an analysis of radio observations of the Lighthouse Nebula with MeerKAT and ASKAP. The excellent sensitivity of MeerKAT on intermediate angular scales reveals highly structured emission from the system and expands our view of the development of the synchrotron nebula downstream of the bow shock of the fast-moving pulsar PSR J1101--6101. For the first time, we detect the extension of the tail to beyond 5pc from the pulsar in the direction opposite to its proper motion, as well as a system of multiple transverse, two-sided emission streaks qualitatively reminiscent of phenomena observed in the vicinity of the Galactic centre region (non-thermal filaments and synchrotron harps). Conversely, we detect no radio counterpart of the long misaligned X-ray jet in the Lighthouse Nebula, suggesting different physical origins for the X-ray and radio features. We also observe an asymmetric distribution of the radio spectral index , with the steepest value located on the side of the misaligned X-ray jet, slightly downstream of its base.

The radio-to-X-ray emission from the tail can be accounted for from a particle population energised close to the pulsar and progressively cooling as it is advected downstream, producing synchrotron emission in a $30-100$\mug magnetic field. The observed broadband synchrotron emission requires about $20-40$\% of the pulsar spin-down power to be converted into relativistic leptons and involves pair multiplicity in the $10^4-10^6$ range, consistent with theoretical expectations and observations of other pulsars. 

We interpret the transverse radio streaks as the occasional release of leptons from the tail into the surrounding medium. The release could be triggered by dynamical instabilities in the tail flow, which would break particle confinement and allow the particles to propagate into the ambient medium. The observed brightening and/or broadening of the tail at the base of the streaks lends support to this scenario and suggests significant reorganisation of the flow. In the Lighthouse Nebula, the observed intensities along the tail and in the streaks suggest that most of the particle content is discharged into the environment within $2-3$\pc (a non-exclusive interpretation is that the magnetic field in the tail undergoes significant decay). Once released, the particles light up the ambient magnetic field and, in the present case, reveal a magnetic field with a coherence length of at least a few parsecs. 
The length and persistence of the streaks imply a degree of magnetic turbulence that is consistent with, or slightly enhanced compared to, the Galactic average at the relevant scales, as estimated from large-scale cosmic-ray transport models. The observed intensity in the streaks does not necessarily call for strong enhancements of the total magnetic field strength, but it is compatible with an ISM magnetic field stronger than the average value ~($3~\mu$G).

We encourage deep radio observations of this and similar BSPWN systems to improve our understanding of non-thermal particle transport in these complex objects. A better understanding of particle escape from pulsar-powered sources is of great importance for several topics, such as the characterisation of magnetic turbulence and transport conditions in the vicinity of cosmic-ray accelerators, the identification of the specific mechanisms leading to the formation of pulsar gamma-ray halos, and the interpretation of the local cosmic-ray positron flux.

\section{Data availability}
The reduced intensity map shown in Fig. \ref{fig:intensity} is available in FITS format at the CDS via \url{http://cdsweb.u-strasbg.fr/cgi-bin/qcat?J/A+A/}.

\begin{acknowledgements}
The MeerKAT telescope is operated by the South African Radio Astronomy Observatory, which is a facility of the National Research Foundation, an agency of the Department of Science, Technology and Innovation.
This scientific work uses data obtained from Inyarrimanha Ilgari Bundara, the CSIRO Murchison Radio-astronomy Observatory. We acknowledge the Wajarri Yamaji People as the Traditional Owners and native title holders of the Observatory site. CSIRO’s ASKAP radio telescope is part of the Australia Telescope National Facility (https://ror.org/05qajvd42). Operation of ASKAP is funded by the Australian Government with support from the National Collaborative Research Infrastructure Strategy. ASKAP uses the resources of the Pawsey Supercomputing Research Centre. Establishment of ASKAP, Inyarrimanha Ilgari Bundara, the CSIRO Murchison Radio-astronomy Observatory and the Pawsey Supercomputing Research Centre are initiatives of the Australian Government, with support from the Government of Western Australia and the Science and Industry Endowment Fund. 
Pierrick Martin, Mickael Coriat, and Alexandre Marcowith acknowledge financial support by ANR through the GAMALO project under reference ANR-19-CE31-0014. 
Barbara Olmi, Elena Amato, Niccolò Bucciantini and Sarah Recchia acknowledge financial support by the European Union – NextGenerationEU RRF M4C2 1.1 grant PRIN-MUR 2022TJW4EJ.
Miroslav Filipovi\'c\ and Sanja Lazarevi\'c acknowledge the Australian Research Council funding through grant DP200100784. 
This work has made use of the SIMBAD database, operated at CDS, Strasbourg, France, and of NASA's Astrophysics Data System Bibliographic Services. The preparation of the figures has made use of the following open-access software tools: Astropy~\citep{Astropy:2013}, Matplotlib~\citep{Hunter:2007}, NumPy~\citep{VanDerWalt:2011}, SciPy~\citep{Virtanen:2020}, and Naima~\citep{zabalza:2015}. The MeerKAT data published here have been reduced using the CARACal pipeline, partially supported by ERC Starting grant number 679627 FORNAX, MAECI Grant Number ZA18GR02, DST-NRF Grant Number 113121 as part of the ISARP Joint Research Scheme, and BMBF project 05A17PC2 for D-MeerKAT. Information about CARACal can be obtained online under the URL: https://caracal.readthedocs.io
\end{acknowledgements}

\bibliographystyle{aa}
\bibliography{aa59134-26.bib}

\label{LastPage}
\end{document}